\documentclass[twocolumn,aps,showpacs,nofootinbib,epsfig]{revtex4}
\usepackage{epsfig}
\usepackage{amsmath}
\usepackage{graphics}
\usepackage{amssymb}
\newcommand{\be}{\begin{equation}}
\newcommand{\ee}{\end{equation}}
\newcommand{\bea}{\begin{eqnarray}}
\newcommand{\eea}{\end{eqnarray}}
\newcommand{\beas}{\begin{eqnarray*}}
\newcommand{\eeas}{\end{eqnarray*}}

\newcommand{\M}{\mathcal{M}}

\begin{document}
\title{Dynamical mass generation in QED with magnetic fields:
arbitrary field strength and coupling constant} 
\author{Eduardo Rojas$^\dagger$, Alejandro Ayala$^\dagger$, Adnan
 Bashir$^{\ddagger}$ and Alfredo Raya$^\ddagger$}    
\affiliation{$^\dagger$Instituto de Ciencias Nucleares, Universidad
Nacional Aut\'onoma de M\'exico, Apartado Postal 70-543, M\'exico
Distrito Federal 04510, M\'exico.\\
$^\ddagger$Instituto de F{\'\i}sica y Matem\'aticas,
Universidad Michoacana de San Nicol\'as de Hidalgo, Apartado Postal
2-82, Morelia, Michoac\'an 58040, M\'exico.}

\begin{abstract}

We study the dynamical generation of masses for fundamental fermions in
quenched quantum electrodynamics, in the presence of magnetic fields of 
{\em arbitrary strength}, by solving the Schwinger-Dyson equation 
(SDE) for the fermion
self-energy in the rainbow approximation. We employ the Ritus eigenfunction
formalism which provides a neat solution to the technical problem of 
summing over all Landau levels. It is well known that
magnetic fields {\em catalyze} the generation of fermion
mass $m$   for arbitrarily small 
values of electromagnetic coupling $\alpha$. For intense fields it is also
well known that $m \propto \sqrt
{eB}$. Our approach allows us to span all regimes of parameters $\alpha$ and
$eB$. We find that 
$m \propto \sqrt {eB}$ provided $\alpha$
is small. However, when $\alpha$ increases beyond the critical value $\alpha_c$
which marks the onslaught of dynamical fermion masses in vacuum, we
find $m \propto \Lambda$, the cut-off required to regularize the ultraviolet
divergences. Our method permits us to verify the results available in 
literature for the limiting cases of $eB$ and $\alpha$. We also point
out the relevance of our work for possible physical applications.

\noindent Preprint Number UMSNH-IFM-F-2008-14.

\end{abstract}


\maketitle

It is well known that fermions can acquire  mass dynamically, 
i.e., through self interactions by means of non perturbative effects without 
the need of a non zero bare mass. 
This phenomenon is called the dynamical mass generation (DMG)
and can be studied in the continuum through the Schwinger-Dyson equations 
(SDEs).  In quantum electrodynamics (QED), DMG takes place only if  the 
electromagnetic coupling $\alpha$ exceeds a critical value $\alpha_c$. 
However, the presence of 
magnetic fields brings about a drastic change and it is possible to generate 
fermion masses for any value of the
coupling. This phenomenon, named {\it magnetic
catalysis}~\cite{Gusynin,Leung,Hong,Ferrer}
has been extensively studied in the situation where the field is strong. 
In this case, only the lowest Landau level 
(LLL) is enough to describe it and the analysis simplifies
considerably. 

Physically, for strong fields, Landau levels are widely spaced making 
it energetically less favorable for virtual particles to populate levels 
other than the LLL. Moreover, for weak coupling, the dynamics describing the 
formation of a condensate is dominated by the LLL. The essence of the
effect is a dimensional reduction brought about by the presence of
the magnetic field that produces an effectively stronger interaction 
among virtual particles and antiparticles in the vacuum. This interaction is
even stronger in the LLL since the only component of the momentum contributing 
to the energy is the longitudinal one and this makes it more easy for virtual
pairs to meet each other and condense. 

Nevertheless, one can envision a situation where the magnetic field is not so 
strong and therefore that low energy levels are close to the LLL in such a way
that virtual particles in those levels also contribute to the dynamics of
condensate formation. Furthermore one could study the phenomenon as a function
of the coupling constant to interpolate between the small coupling domain
where the dynamics is magnetic field driven to the strong coupling domain
where the dominance should switch to the description of dynamical symmetry
breaking in vacuum. 

To the best of our knowledge, calculations aiming to describe magnetic catalysis
have non been performed considering the contribution of all Landau levels, 
for arbitrary field strength and coupling constant, perhaps because such 
calculations involve the technical challenge of carrying out a seemingly
prohibitive sum over these levels. In this work, we undertake 
the study of DMG in QED in the
rainbow  approximation in the presence of magnetic fields of arbitrary
intensity. We present a solution to the above technical difficulty and use it
to carry out a detailed quantitative analysis of the dynamically generated
fermion mass for arbitrary values of the coupling constant and magnetic field
intensities. 

It has been shown~\cite{Ritus} (see also Ref.~\cite{Incera1}) that the
mass operator in the  presence of an electromagnetic field can be written as a
combination of the structures
$\gamma^\mu\Pi_\mu$, $\sigma^{\mu\nu}F_{\mu\nu}$, $
(F_{\mu\nu}\Pi^\nu)^2$, $\gamma_5F_{\mu\nu}\tilde{F}^{\mu\nu}
   \label{structures}
$
which commute with the operator $(\gamma_{\mu} \Pi^{\mu})^2$, where 
$
   \Pi_\mu = i\partial_\mu - eA^{\mbox{\tiny{ext}}}_\mu,\
   F_{\mu\nu}=\partial_\mu A^{\mbox{\tiny{ext}}}_\nu -
              \partial_\nu A^{\mbox{\tiny{ext}}}_\mu,\
   \tilde{F}^{\mu\nu}=\frac{1}{2}\epsilon^{\mu\nu\lambda\tau}
   F_{\lambda\tau},\
   \sigma_{\mu\nu}=\frac{i}{2}[\gamma_\mu , \gamma_\nu]
$
and $A^{\mbox{\tiny{ext}}}$ is the external vector
potential. We take 
$A_{\mbox{\tiny{ext}}}^{\mu}=B(0,-y/2,x/2,0)$ in such a way that it
gives rise to a constant magnetic field
${\mathbf{B}}=B{\mathbf{\hat{z}}}$.
The two-point fermion Green's function $G(x,y)$ is found through the equation
\bea
   \gamma\cdot\Pi(x)G(x,y)\!-\!\!\int \!\!d^4x'M(x,x')G(x',y)
   =\delta^4(x-y) \;,
   \label{Greeneq}
\eea
where the mass operator $M(x,y)$
is described by its SDE in the rainbow approximation as 
\bea
   M(x,x')=-ie^2\gamma^\mu G(x,x')\gamma^\nu D^{(0)}_{\mu\nu}(x-x').
   \label{SDcoord}
\eea 
In the presence  of  a constant
external field,  the fermion  asymptotic  states  are no   longer  free
particle  states, but  instead are described by eigenfunctions of the operator
$(\gamma^{\mu}\Pi_{\mu})^{2}$. Therefore,  it is  convenient  to  work  in  the
representation  spanned  by  these eigenfunctions, rendering  the  mass
operator diagonal and allowing to write the equation for the mass function
as~\cite{ayala},
\begin{widetext}
\begin{align}
   {\mathcal{M}}(p_\parallel,n_p)=-ie^2\sum_{\sigma_k,\sigma_p=\pm
   1} 
   \sum_{n_k,s_k}\frac{s_k! (n_p-\frac{(\sigma_p+1)}{2})!}{s_p!
   (n_k-\frac{(\sigma_k+1)}{2})!} 
   \int \frac{d^4q}{(2\pi)^4}\frac{e^{-q_\perp^2/2\gamma}}{q^2+i\epsilon}
   \frac{{\mathcal{M}}((p-q)_\parallel,n_k)}{(p-q)_\parallel^2-
   2eBn_k-{\mathcal{M}}^2((p-q)_\parallel,n_k)}
   \nonumber\\
   \times\left(\frac{q_\perp^2}{4\gamma}\right)^
   {l_k-l_p-\frac{(\sigma_k-\sigma_p)}{2}}\!\!
   \left[ 2 + \frac{(1-\xi)}{q^2}\left(q_\perp^2(1-\delta_{\sigma_p\sigma_k})
   -q_\parallel^2\delta_{\sigma_p\sigma_k}\right)\right]\!\! 
   \left[L_{n_p-\frac{(\sigma_p+1)}{2}}^{n_k-n_p-
   \frac{(\sigma_k-\sigma_p)}{2}} 
   \left( \frac{q_\perp^2}{4\gamma} \right)  \right]^2\!\!
   \left[L_{s_k}^{s_p-s_k} \left( \frac{q_\perp^2}{4\gamma} 
\right)\right]^2\, ,
\end{align}
\end{widetext}
where $q_\perp = (0,q_1,q_2,0)$,  $q_\parallel=(q_0,0,0,q_3)$, and thus, in
Minkowski space, $q^2=q_\parallel^2-q_\perp^2$. Also,  $\gamma=eB/2$,
$p^2=E_p^2-p_z^2-2eBn$, $\xi$ is the covariant gauge parameter and 
$L_{n}^{m}$ are  Laguerre  functions.
We shall assume that the wave function
renormalization equals one. 
After the structures upon which the mass operator $M$ depends have been
accounted for, the mass function should be a scalar matrix whose components
can in principle be different in the transverse and longitudinal directions
due to the presence of the field. Here, let us work with the ansatz that
${\mathcal{M}}$ is proportional to the unit matrix. The assumption is good
when considering small momentum~\cite{Leung}.
 
We expect that ${\mathcal{M}}((p-q)_\parallel,n_k)$ should be 
independent of $s_k$ since the energy only depends on the
principal quantum number $n_k$. We also assume that
${\mathcal{M}}((p-q)_\parallel,n_k)$ is a slowly varying function of
$n_k$ and thus make the approximation
${\mathcal{M}}((p-q)_\parallel,n_k)\sim{\mathcal{M}}((p-q)_\parallel,n_k=0)$.
For consistency we consider the case $n_p=0$. Hereafter, we employ the
more convenient notation
${\mathcal{M}}(k_\parallel,n_k=0)\equiv{\mathcal{M}}(k_\parallel)$ 
for generic arguments of the mass function. With these
considerations the sum over $s_k$ can be performed
\begin{align}
\M(p_{\parallel})=&-ie^2\sum_{\sigma_{p},\sigma_{k}}\sum_{k=0}^{\infty}\int
\frac{d^4q}{(2\pi)^4}\frac{e^{-\frac{q_{\perp}^2}{4\gamma}}}{q^2+i\varepsilon}
\nonumber\\  
& \hspace{-1.2cm} \times \frac{\M((p-q)_\parallel)}{\left\{(q-p)_\parallel^2-
2eB(k+(\sigma_{k}+1)/2)\right\}-\M^2 \left((p-q)_\parallel\right)}\nonumber\\  
& \hspace{-1.2cm} \times 
\left\{2+(1-\xi)(1-\delta_{\sigma_{p}\sigma_{k}})\frac{q_{\perp}^2}{q^2}
-(1-\xi)\delta_{\sigma_{p}\sigma_{k}}\frac{q_{\parallel}^2}{q^2}\right\}
\nonumber\\    
& \hspace{-1.2cm} \times 
(-1)^{k-m}L_{m}^{k-m} \left( \frac{q_{\perp}^2}{4\gamma} \right)
L_{k}^{-(k-m)} \left( \frac{q_{\perp}^2}{4\gamma} \right), 
\end{align}
where $k=n_{k}-\frac{\sigma_k+1}{2}$. It is worth mentioning that after
summing over $s_k$, the resulting equation is the same as Eq.~(50) in
Ref.~\cite{Leung} when considering  $n_k=0$. It corresponds to the
strong field limit. Using the identity
\bea
   && \hspace{-0.5cm} \frac{1}{(p-q)_\parallel^2-2eB(k+\frac{\sigma_k+1}{2}) -
   {\mathcal{M}}^2((p-q)_\parallel)}=\nonumber\\
   && \hspace{-0.5cm} -i\int_{0}^{\infty}ds
   e^{is[(p-q)_\parallel^2-2eB(k+\frac{\sigma_k+1}{2}) -
   {\mathcal{M}}^2((p-q)_\parallel)+i\epsilon]} \;,
   \label{denomitor}
\eea 
and carrying out the sums over $k$, 
$\sigma_k$ and $\sigma_p$, we get
\begin{align}
   \M(p_\parallel)=&-e^2\int\frac{d^4Q}{(2\pi)^4}
   \frac{\M \left( (\frac{p}{4\gamma}-Q)_\parallel \right)}{Q^2+i\varepsilon}
   \int_{0}^{\infty}d\tau
   \nonumber\\
 &\hspace{-1.3cm}  
\times \ e^{-Q_{\perp}^2\left[1-\exp\left(-i\tau\right)\right]} 
   e^{i\left[(2\sqrt{\gamma}Q-p)_\parallel^2
   -\M^2\left( (\frac{p}{4\gamma}-Q)_\parallel \right)
+i\epsilon\right]\frac{\tau}{4\gamma}}\nonumber\\
 &\hspace{-1.3cm}
   \times \Biggl[\left\{2-(1-\xi)\frac{Q_{\parallel}^2}{Q^2}\right\}
    +\left\{2+(1-\xi)\frac{Q_{\perp}^2}{Q^2}\right\} e^{-i \tau}  
\Biggl],
   \label{minkowsky}  
\end{align}
where $Q=\frac{q}{2\sqrt{\gamma}}$ and $\tau=4\gamma s$. This is the
integral equation for the mass function in Minkowski space. To perform the
integral in the variable $\tau$ we notice that the integrand dies out
sufficiently rapidly for large imaginary values of $\tau$. We can thus close
the contour of integration on a path whose first leg is a horizontal line
along the real $\tau$-axis, continued along the quarter-circle at infinity in
the right-lower quadrant and finally along the negative imaginary
$\tau$-axis. Using Cauchy's theorem, we can thus write Eq.~(\ref{minkowsky}) as
\begin{align}
   \M(p_\parallel)=&e^2\int\frac{d^4Q}{(2\pi)^4}
   \frac{\M ((\frac{p}{4\gamma}-Q)_\parallel)}{Q^2}
   \int_{0}^{\infty}d\tau\nonumber\\
 &\hspace{-1.3cm}  \times\ 
   e^{-Q_{\perp}^2\left[1-\exp\left(-\tau\right)\right]}
   e^{-\left[(2\sqrt{\gamma}Q-p)_\parallel^2 
   +\M^2\left(
   (\frac{p}{4\gamma}-Q)_\parallel \right)\right]\frac{\tau}{4\gamma}}
  \nonumber\\
 &\hspace{-1.3cm}
   \times \Biggl[\left\{ 2-(1-\xi)\frac{Q_{\parallel}^2}{Q^2}\right\}
   +\left\{2-(1-\xi)\frac{Q_{\perp}^2}{Q^2}\right\} e^{-\tau} \Biggl].
   \label{minkowskytoeucledian}
\end{align}
To guaranty convergence of the integral over $\tau$, we need to consider
momenta $Q$ and $p$ in Euclidian space and accordingly, a Wick rotation on $Q$
has already been performed in Eq.~(\ref{minkowskytoeucledian}). 
We now perform the change of variable
$x=e^{-\tau}$ to get 
\begin{align}
   \M (p_\parallel)=& e^2\int\frac{d^4Q}{(2\pi)^4}
   \frac{\M ((\frac{p}{4\gamma}-Q)_\parallel)} {Q^2}\int_{0}^{1}
   dxe^{-Q_{\perp}^2\left(1-x\right)}\nonumber\\ 
 &\hspace{-1cm}  \times \ x^{\left[(Q-p)_\parallel^2
   +\frac{\M^2}{2eB}-1\right]}
   \nonumber\\ 
 &\hspace{-1cm}  \times 
\Biggl[\left\{2-(1-\xi)\frac{Q_{\parallel}^2}{Q^2}\right\}
   +\left\{2-(1-\xi)\frac{Q_{\perp}^2}{Q^2}\right\}x\Biggl],
   \label{inx} \;
\end{align}
where we have not shown the argument of $\M^2$ to avoid cumbersome 
notation.

\begin{figure}[t!] 

\epsfig{file=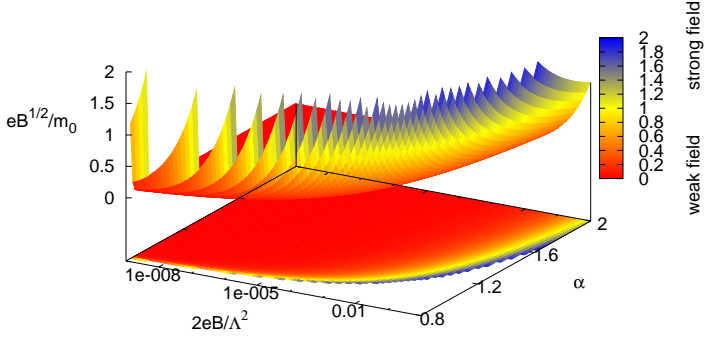,width=0.7\columnwidth,angle=-90}

\caption{(Color online) Relative strength of the magnetic field
$\sqrt{eB}/m_0$,  
$m_0$ is the dynamical mass in the vacuum, as a function 
of the coupling $\alpha$ and the parameter $2eB/\Lambda^2$. Studying this 
dependence is possible
only for $\alpha > \alpha_c=\pi/4 $ where $m_0 \neq 0$. 
Towards yellow and blue, magnetic field becomes strong and towards red,
it becomes weak. Note that for higher couplings, it is easier to access 
weak fields. 
On the other hand, lower couplings imply strong magnetic fields unless 
$eB << \Lambda^2$.}
\label{fig1}
\end{figure}

To illustrate the calculation of the mass function, we work in the
Feynman gauge, $\xi=1$. It is well known that in the ladder approximation, the
most preferable gauge is the Landau gauge $\xi=0$ because it comes the closest
to satisfying the Ward-Takahashi identity. However, within the ladder
approximation, working with a different value
of $\xi$ gets reflected in the fact that--in the
free field case-- the coefficient of the coupling constant changes
from $1$ to $(1+\xi/3)$. Therefore, in the ladder approximation, the
physical picture of dynamical mass generation should be similar in all gauges
close to the Landau gauge, including $\xi=1$. For example, the critical
coupling changes from approximately $1.04$ for Landau gauge to $0.78$ for
the Feynman gauge (see fig.~\ref{fig4}). The implicit assumption is that
the same happens in the presence of the field. This can be entirely justified
for the weak field limit since one always keeps close to vacuum. In the strong
field case it is also justified, since our results reproduce this very well
studied case (see for instance Ref.~\cite{Leung} where the analysis was also
carried out in the Feynman gauge).  

Upon the change of variables
\mbox{$Q\rightarrow Q+p_\parallel /\sqrt{4\gamma}$},   
\begin{align}
   \M\left(p_\parallel /\sqrt{4\gamma}\right)=&\ e^2\int\frac{d^4Q}{(2\pi)^4}
   \frac{\M (Q_\parallel)}{(Q+p_\parallel )^{2}}\notag\\ 
   \times&\int_{0}^{1}dxx^{\lambda}e^{-Q_\perp^2(1-x)}\ 2(1+x),
   \label{efeynman}
\end{align}
where 
\mbox{$\lambda=Q_\parallel^{2}+\frac{\M^{2}(Q_\parallel)}{4\gamma}-1$}.
To perform the angular integration, we write
\mbox{$d^4Q=d^2Q_{\perp}d^2Q_{\parallel}=\frac{\pi}{2}dQ_\perp^2
dQ_\parallel^2d\theta$}, where $\theta$ is the angle between $Q_\parallel$ and
$p_\parallel$. Hereafter, we assume that the mass function depends only on the
magnitude of its argument. The angular integration is now easily carried out
and the result can be expressed as
\begin{align}
   \mathcal{M}(p_\parallel/\sqrt{4\gamma})=
   &\frac{e^2}{2(2\pi)^2}\int_0^1{dx[1+x]}
   \int dQ_\parallel^2x^{\lambda}\mathcal{M}(Q_\parallel)\nonumber\\  
 & \hspace{-2cm} \times\int_0^\infty 
  \frac{dQ_\perp^2 \; e^{-Q_\perp^2(1-x)}}
   {\sqrt{\left[Q_\parallel^2 
   -\frac{p_\parallel^2}{4\gamma}\right]^2
   +2Q_\perp^2\left[\frac{p_\parallel^2}{4\gamma}
   +Q_\parallel^2\right]+Q_\perp^4}}.
\end{align}
We now approximate the argument of the square root in the
denominator by intervals. For $p_\parallel^2/4\gamma\ge Q_\parallel^2$
we take $\frac{p_\parallel^2}{4\gamma}+Q_\parallel^2\sim
\frac{p_\parallel^2}{4\gamma}$. Conversely, for $Q_\parallel^2\ge
p_\parallel^2/4\gamma$ we take $\frac{p_\parallel^2}{4\gamma}+
Q_\parallel^2\sim Q_\parallel^2$. With this approximation, the
integral over $Q_\perp$ can be analytically performed and the result is
\begin{align}
   \mathcal{M}(p_\parallel/\sqrt{4\gamma})=\frac{e^2}{2(2\pi)^2}
   \int_0^1{dx[1+x]x^\lambda 
   \mathcal{M}(Q_\parallel)}\notag\\ 
   \times\Biggl\{\int_0^\frac{p^2}{4\gamma}dQ_\parallel^2 
   \exp{\left[(1-x)\frac{p^2}{4\gamma}\right]}
   \Gamma\left[0,(1-x)\frac{p^2}{4\gamma}\right]\nonumber\\
   +\int_{p^2/4\gamma}^\infty{dQ_\parallel^2}
   \exp{\left[(1-x)Q_\parallel^2\right]}
 \Gamma\left[0,(1-x)Q_\parallel^2\right]
   \Biggr\},
   \label{lsst}
\end{align}
where $\Gamma(x,y)$ is the incomplete gamma function. 
Notice that the approximations leading to Eq.~(\ref{lsst}) 
make no reference to the strength of the magnetic
field. Therefore, this equation is valid for
arbitrary magnetic field intensities. In the following, we use the result in
Eq.~(\ref{lsst}) to numerically explore the 
behavior of the mass function when varying either the magnetic field
strength or the value of the coupling constant. The validity of our
approximations gets a numerical affirmation when in certain limiting 
parametric regime of $\alpha$ and $eB$, we retrieve the results already
known in the literature.
\begin{figure}[t!] 
 \epsfig{file=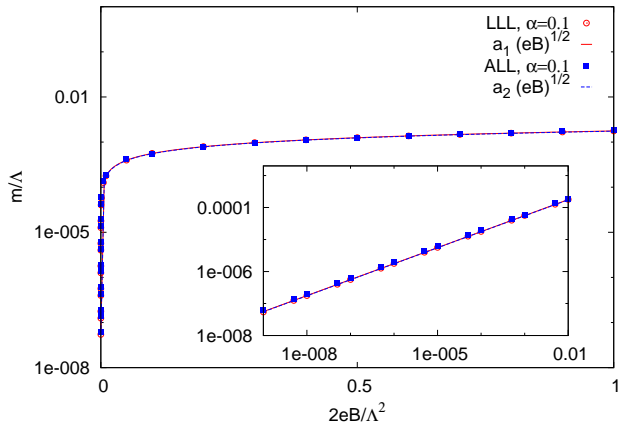,width=0.7\columnwidth,angle=-90}

\caption{(Color online) Dynamical mass as a function of the magnetic field in
units of ultraviolet cut-off in the weak coupling regime. Blue dots 
correspond to summing over all Landau levels whereas red dots
correspond to the LLL approximation. We have used $\alpha=0.1$. 
Furthermore, $a_1=0.00176653$ and 
$a_2=0.00174613$. As $\alpha< \alpha_c$, we get $m \propto \sqrt {eB}$,
no matter the strength of the magnetic field.}
\label{fig2}
\end{figure}

\begin{figure}[t!] 
 \epsfig{file=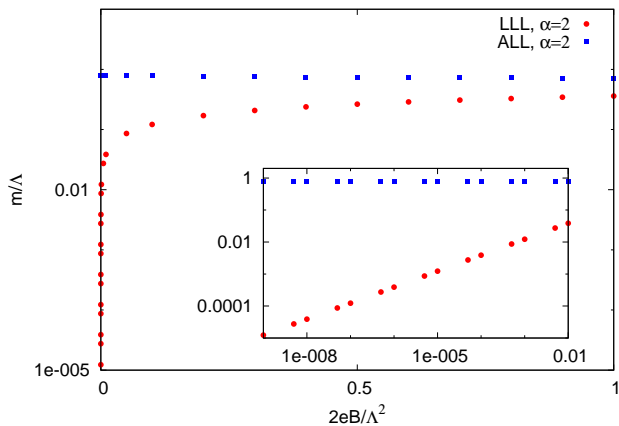,width=0.7\columnwidth,angle=-90}

\caption{(Color online) Dynamical mass as a function of the magnetic field in
units of ultraviolet cut-off in the strong coupling regime. Blue dots 
correspond to
  summing over all Landau levels whereas red dots correspond to the LLL
  approximation. We have used $\alpha=2$. As we practically get a flat line,
we conclude that $m \propto \Lambda$. }
\label{fig3}
\end{figure}

For $\alpha > \alpha_c (=\pi/4)$, we determine the strength of the magnetic 
field  by comparing it with the
dynamically generated mass $m_0$ in the vacuum. $\sqrt{eB}/m_0 >>1$ is the
strong field limit whereas  $\sqrt{eB}/m_0 << 1$ corresponds to the weak
field limit. In the Feynman gauge, this ratio has been depicted in fig.~1.
For $\alpha < \alpha_c$, $m_0=0$. Therefore, the strength of the magnetic field
can only be compared with the ultraviolet cut-off $\Lambda$.

Therefore, we first study the dependence of the mass function on the magnetic 
field strength for $\alpha < \alpha_c$. We concentrate on the dependence of
the dynamically generated mass $m$, defined as the mass function evaluated at
zero momentum.
In fig.~2, we plot the solution to 
Eq.~(\ref{lsst}) containing the sum over all Landau levels
for a very small value of the coupling,
$\alpha=0.1$, as a function of the magnetic field strength, ranging from
$eB/\Lambda^2$ as large as 1 to as low as $10^{-9}$. Our results are 
virtually the same as obtained from the LLL approximation~\cite{Leung}
as expected.

\begin{figure}[t!] 
 \epsfig{file=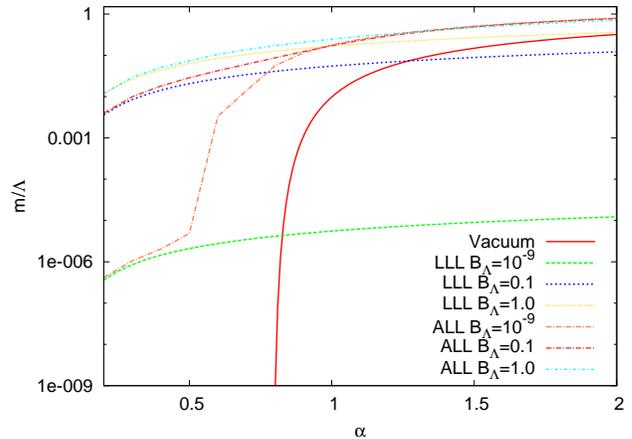,width=0.7\columnwidth,angle=-90}

\caption{(Color online) Dynamical mass as a function of the coupling constant
  for different values of the magnetic field. We have used the notation
  ${\rm B}_{\Lambda}=2 e B/{\Lambda^2}$. For
  comparison, we show  the behavior of the dynamical mass in the
  absence of a magnetic field. Note that for large values of ${\rm
  B}_{\Lambda}$ the dynamical mass obtained by considering just the
  contribution of the LLL is practically the same as the one obtained by
  considering the contribution of all Landau levels. The situation changes
  for small values of ${\rm B}_{\Lambda}$ where it can be seen that other
  levels than just the LLL contribute to the dynamical mass.}
\label{fig4}
\end{figure}
We now turn our attention to the study of the mass function for
$\alpha > \alpha_c$. Corresponding results are shown in fig.~3.  
Note that considering all Landau levels changes the results
considerably in comparison with the predictions of the LLL. The
dynamically generated mass $m \propto \Lambda$. Note how the 
proportionality of the mass function switches from $\sqrt {eB}$
to $\Lambda$ when we move from $\alpha < \alpha_c$ to $\alpha > \alpha_c$.

Finally, we present the 
explicit dependence of $m$ on the coupling
constant in fig.~4 both for the LLL and all Landau levels contributing.
For comparison, we show the corresponding dynamically generated
mass in vacuum. Notice that, as is known, magnetic field of 
{\em arbitrary strengths, however small} ($eB << \Lambda^2$) catalyze the
appearance of a dynamically generated mass for any value of $\alpha$. For
large values of ${\cal B}_{\Lambda}=2 e B/{\Lambda^2}$, the dynamical mass
obtained by considering just the contribution of the LLL matches onto the one
obtained by considering the contribution of all Landau 
levels. Going towards decreasing $\alpha$, this matching gets triggered early
on if $eB \approx \Lambda^2$. However, one has to go to very small values of
$\alpha$ to achieve the same if $eB << \Lambda^2$ as we might expect
on physical grounds. The situation changes for small values of ${\cal
B}_{\Lambda}$ where it can be seen that other levels than just the LLL
contribute to the dynamical mass. In this last case, the transition region
is around the critical value
of the coupling constant in vacuum, $\alpha_c$. For $\alpha >\alpha_c$ the
dynamics is dominated by the strength of the coupling constant, as the largest
contribution to the mass comes from the intensity of self-interactions rather
than from the magnetic field contribution.  

Notice how the results depicted in fig.~4 are harmoniously consistent with
those in figs.~2 and 3. Let us first focus on fig. 2. It has been
drawn for $\alpha=0.1$. This value of $\alpha$ corresponds to the far left of
fig.~4 where our results including all the Landau levels are practically
the same as the LLL results, i.e., the dynamically generated mass is
proportional to $\sqrt{eB}$. Let us now compare fig.~4 with fig.~3. Figure~3
has been obtained by setting 
$\alpha=2$, which corresponds to the far right of fig.~4. Note that at that
end, our results including all Landau levels match onto each other for
all values of the parameter ${\cal B}_{\Lambda}$ ranging from $10^{-9}$ to
$1.0$. This implies that for large values of $\alpha$, $m/\Lambda$ is in fact
independent of ${\cal B}_{\Lambda}$. Therefore, we conclude that
$m \propto \Lambda$, the result we earlier deduced from fig.~3. This shows
how, fig.~4 is consistent both with figs.~2 and~3. In fact, it is
complimentary. It clearly shows that for intermediate regions of $\alpha$, we
must take into account all Landau levels for a quantitatively correct
description of magnetic catalysis. 

In conclusion, 
we have studied the phenomenon of magnetic catalysis for arbitrary
values of the coupling constant and magnetic field strength
by taking into account the contribution 
from all the energy levels in the system. We have also shown 
that the phenomenon is {\em not always} restricted to the contribution of 
the LLL. This is especially evident for low and moderate 
values of the magnetic field. In the special case of the
strong field limit and for small values of the coupling constant, 
studied extensively in the literature, our 
analysis confirms the well known behavior of the dynamically 
generated mass. 

Our general results can have interesting 
cosmological consequences, as is exemplified for 
instance, in the study of the electroweak phase transition. 
Recall that this transition took place during the early universe 
for temperatures around $T=100$ GeV. It has been recently shown 
that within the standard model, if a primordial magnetic field 
was present at this epoch, the phase transition becomes 
stronger first order~\cite{ayala1}. The analysis is based on 
the study of the effective potential whose development with 
temperature is driven by the Higgs condensate. However a fermion mass
generated by the magnetic field is 
tantamount of a fermion condensate whose contribution to the 
development of the phase transition, and in particular, to 
the true vacuum expectation value of the Higgs field could 
further modify the nature of the transition and influence 
the baryogenesis scenario. 

Another interesting question concerns the study of higher order
effects. It is well known that such effects change the behavior of the
dynamically generated mass as a function of the coupling constant for strong
fields and weak coupling (see the second to the last of
Refs.~\cite{Gusynin}). In the context of the present work it is possible to
study how this change evolves as the magnetic field and coupling constant take
on arbitrary values. These issues deserve further investigation that we are
currently pursuing and will be reported elsewhere.

The authors acknowledge useful comments and suggestions by V. de la Incera,
E. Ferrer and V. Gusynin. Support for this work has been received in part by
SNI as well as COECyT, CIC, CONACyT and PAPIIT grants under  
projects CB070218-4, CB0702162-4, 4.12, 4.22, 46614-I, 40025-F and IN116008,
respectively. E.R. thanks the financial support of DGEP-UNAM.

\end{document}